# Materealize: a multi-agent deliberation system for end-to-end material design and synthesis

Seongmin Kim[1†], Jaehwan Choi[1†], Kunik Jang[1†], Junkil Park[1,2], Varinia Bernales[3,4,7], Alán Aspuru-Guzik[3,4,5,6,7,8,9,10], and Yousung Jung[1,2*]

[1] *Department of Chemical and Biological Engineering (BK21 four), and Institute of Chemical Processes, Seoul National University, 1 Gwanak-ro, Gwanak-gu, Seoul 08826, Korea*
[2] *Institute of Engineering Research, Seoul National University, 1 Gwanak-ro, Gwanka-gu, Seoul 08826, Korea*
[3] *Department of Chemistry, University of Toronto, 80 St. George St., Toronto, ON M5S 3H6, Canada*
[4] *Department of Computer Science, University of Toronto, 40 St George St., Toronto, ON M5S 2E4, Canada*
[5] *Department of Materials Science & Engineering, University of Toronto, 184 College St., Toronto, ON M5S 3E4, Canada*
[6] *Department of Chemical Engineering & Applied Chemistry, University of Toronto, 200 College St., Toronto, ON M5S 3E5, Canada*
[7] *Acceleration Consortium, 700 University Ave., Toronto, ON M7A 2S4, Canada*
[8] *Vector Institute for Artificial Intelligence, W1140-108 College St., Schwartz Reisman Innovation Campus, Toronto, ON M5G 0C6, Canada*
[9] *Canadian Institute for Advanced Research (CIFAR), 661 University Ave., Toronto, ON M5G 1M1, Canada*
[10] *NVIDIA Corporation, Santa Clara, CA, USA*

† These authors contributed equally to this work.

* Email: yousung.jung@snu.ac.kr

## Abstract

We propose *Materealize*, a multi-agent system for end-to-end inorganic materials design and synthesis that orchestrates core domain tools spanning structure generation, property prediction, synthesizability prediction, and synthesis planning within a single unified framework. Through a natural-language interface, *Materealize* enables non-experts to access computational materials workflows and obtain experimentally actionable outputs for material realization. *Materealize* provides two complementary modes. In *instant mode*, the system rapidly composes connected tools to solve diverse inorganic tasks—including property-conditioned synthesizable candidate design with synthesis recipes, diagnosis, and redesign of unsynthesizable structures, and synthesizable data augmentation—within a few minutes. In *thinking mode*, *Materealize* applies multi-agent debate to deliver more refined and information-rich synthesis recommendations, including reasoning- and model-driven synthesis routes and mechanistic hypotheses. The mechanistic hypotheses are validated by direct comparison with the literature for known mechanisms and further supported by physics-grounded simulations for novel synthesis pathways. By combining tool-level accuracy with reasoning-level integration, *Materealize* can bridge the gap between computational discovery and practical experimental realization.

## Introduction

Advances in computational chemistry and machine learning have transformed the landscape of computationally guided materials discovery.[1-3] A wide range of tasks in material chemistry—ranging from quantum chemical calculations to chemical property predictions—are increasingly being developed by artificial intelligence (AI) models.[4] Especially for material discovery, a variety of machine learning (ML) models have been developed to facilitate material property prediction and inverse design toward desired properties, including bandgap, magnetic density, and crystal density.[5-7] In parallel, to support the experimental realization of these promising candidates with desired properties, stability, and synthesizability of inorganic crystals have also been studied for developing thermodynamics-based or data-driven formability predictors.[8,9] These tools have enabled researchers to explore candidates that are not only high-performing but also experimentally feasible. Moreover, several synthesis-prediction models can also suggest concrete synthesis pathways and parameters, providing specific guidance on how to perform the reaction, which could drastically accelerate material realization.[10-13]

Despite these advancements, a significant gap remains between computational-guided materials and their practical experimental realization. One major barrier is that computational chemistry tools and ML models have been developed in a fragmented, task-specific set of tools, without a unified framework that seamlessly connects aspects of design, evaluation, and synthesis. As a result, even when powerful tools exist, navigating them often requires substantial domain expertise. This complexity limits accessibility for lots of experimental chemists and materials scientists who may not be proficient in computational software or in adopting ML workflows.

The recent emergence of large language models (LLMs) and demonstrations of their effectiveness in materials chemistry have begun to bridge this gap.[14] LLM-based interfaces allow non-experts to interact with complex tools using natural language, bypassing the need for complex model configuration, hyperparameter tuning, or specialized ML knowledge.[15-24] Furthermore, agentic AI approaches—where LLMs autonomously orchestrate multiple computational and ML tools—have shown that it is possible to perform complex domain-specific tasks by delegating tool use, decision-making, and multi-step reasoning to the model itself.[25-38] Importantly, grounding agent actions in external tools (e.g., databases, simulators, and ML tools) can drastically reduce hallucinations by enforcing evidence-based outputs and validation.[26,39] These developments offer a pathway to unifying previously isolated computational tools into coherent, end-to-end workflows, while also enabling practical use by non-experts who may not be familiar with computational methods.

In this work, we propose *Materealize*, a unified multi-agent system that enables end-to-end inorganic materials design and synthesis planning for material realization. *Materealize* integrates structure generation, synthesizability assessment, and synthesis method prediction into a single agentic framework. Unlike other existing inorganic design agents,[33-36] *Materealize* extends beyond "*materials design*" to incorporate accurate "*synthesis planning*" and "*mechanistic insight*", thereby offering more actionable guidance for experimentalists. The system operates in two complementary modes: *instant mode* and *thinking mode*. In *instant mode*, users can express target properties in natural language (e.g., "I want to design an inorganic material with a bandgap around 1.34 eV."), and *Materealize* automatically invokes the necessary tools—including conditional generative models, synthesizability predictors, and synthesis method models—to propose synthesizable candidates along with plausible synthesis procedures within approximately 1–2 minutes per material. Moreover, because the *Materealize* instant mode integrates several additional ML tools—including explanation and redesign models—it can support a wide range of downstream inorganic materials tasks, such as diagnosing and fixing unsynthesizable structures or augmenting synthesizable data.

However, other existing agentic systems typically rely on a passive "tool-calling" paradigm: the accuracy of the final result is ultimately bounded by the accuracy of predefined tools. To address this limitation, *Materealize* also incorporates a *thinking mode*, where we deploy a multi-agent debate (MAD) approach[32] to enable reasoning-driven synthesis planning. In this mode, each agent is grounded in a distinct ML tool and acts as a critical evaluator of synthesis pathways proposed by heterogeneous tools. The agents collaboratively critique, cross-check, and refine candidate routes, integrating multi-modal evidence to produce a final recommendation that is both more accurate and chemically plausible. Notably, the MAD framework also allows *Materealize* to propose hypothesized synthesis mechanisms, marking one of the first attempts to extract mechanistic insight from agentic systems in inorganic chemistry.

Together, these capabilities position *Materealize* as a step toward autonomous, synthesis-aware materials design—offering both accessibility for non-experts and advanced reasoning for expert-level investigation. Moreover, *Materealize* is designed to be highly extensible, making it easy to integrate new tools and seamlessly expand its functionality. We provide a web-based demo of *Materealize* to broaden its practical usability (https://huggingface.co/spaces/SNU-MICC/Materealize).

## Results and Discussion

### Overall architecture of Materealize

The architecture of *Materealize* comprises two complementary operation modes, an *instant mode* and a *thinking mode*, which together enable both rapid task execution and deeper, reasoning-driven synthesis analysis (**Fig. 1**). At a high level, the instant mode is designed for fast, tool-centric workflows, whereas the thinking mode is designed for different tool-grounded multi-agent scientific reasoning about synthesis routes and mechanisms.

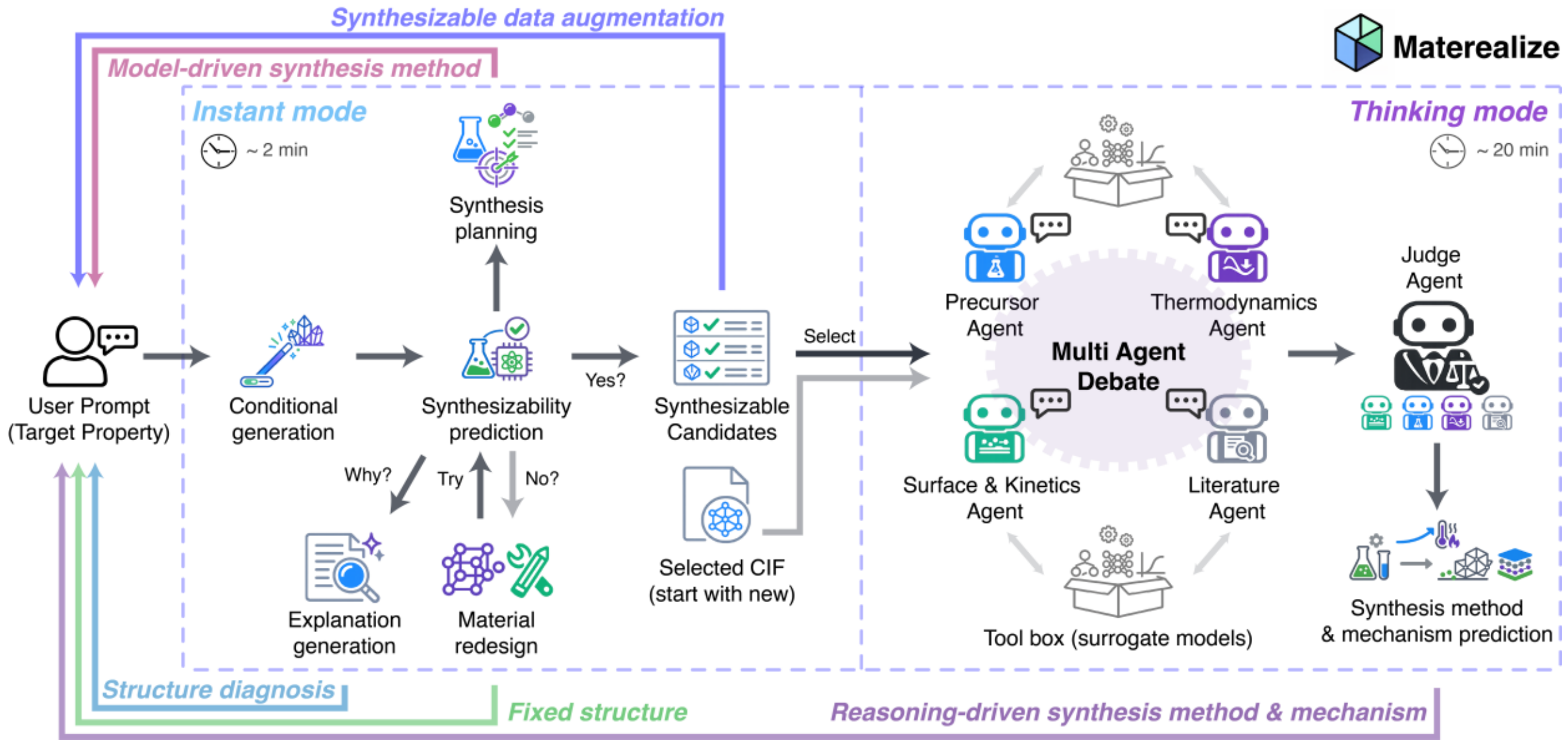

**Fig. 1 | Overall architecture of *Materealize*.** *Materealize* comprises two operation modes; *Instant mode* uses direct tool execution to rapidly address diverse inorganic design and synthesis tasks, making it well-suited for interactive exploration and large-scale screening. In contrast, *thinking mode* applies multi-agent debate reasoning to deliver richer, mechanistically grounded synthesis recommendations for complex or high-stakes decisions.

In ***instant mode***, *Materealize* connects a suite of specialized tools that cover generative design, synthesizability prediction, synthesis recipe prediction, structural diagnosis and redesign, and data handling:

I. *ChargeDIFF* and *Chemeleon* - inorganic structure generative models[7,40] capable of both unconditional crystal generation and conditional generation with respect to target properties such as bandgap, magnetic moment density, crystal density, and chemical composition.

II. *PU-CGCNN* and *SynCry* - inorganic synthesizability predictors based on two complementary

representations: a crystal-graph neural network (PU-CGCNN)[41] and a fine-tuned LLM-based model operating on CIF-string representations (SynCry)[42].

III. *ALIGNN* - a model for predicting various material properties [43] based on the graph neural network.

IV. *ElemwiseRetro* - a data-driven inorganic synthesis recipe predictor[13] that suggests precursors and heating temperatures, implemented as a Roost-based ML model on compositional features.

V. *StructGPT* - an LLM-based inorganic explanation model[44] that diagnoses why a given structure may be difficult/possible to synthesize and provides human-readable rationales.

VI. *SynCry* - an inorganic structure-redesign model[42] that converts unsynthesizable or low-feasibility structures into more synthesizable alternatives.

VII. *Data I/O* and *Representation tools* - utilities for handling inorganic data formats and representations, including CIF file loading/saving, pymatgen-based formula parsing,[45] crystal-graph construction,[5] CIF-string conversion,[42] and Robocrystallographer-based descriptive text generation[46] for crystal structures.

By orchestrating these tools, instant mode can address a variety of inorganic design tasks within a single unified workflow. Typical tasks include (i) target-property-conditioned design of synthesizable candidates with corresponding synthesis methods, (ii) diagnosis and repair of unsynthesizable structures, and (iii) unconditional generation of synthesizable candidates for data augmentation. Users can initiate these workflows with a simple natural-language query (e.g., "I want to design an inorganic material with a bandgap around 1.34 eV and obtain a detailed, step-by-step synthesis guide."), after which *Materealize* automatically selects and composes the appropriate tools. In practice, this enables end-to-end inorganic design and synthesis suggestion within only a few minutes, making the system suitable for interactive use.

While instant mode already provides an integrated pipeline that connects generative models, property predictors, synthesizability predictors, and synthesis planners, it remains fundamentally constrained by the limitations and biases of the individual tools. The agent system's output, like many other agentic systems, is inevitably dependent on the accuracy of its underlying models and is further limited by the absence of explicit cross-checking or higher-level scientific reasoning.

To overcome these limitations, *Materealize* provides a ***thinking mode*** that deploys a multi-agent debate (MAD) framework.[32] In this mode, four domain-specialized agents—*Precursor*, *Thermodynamics*, *Surface & Kinetics*, and *Literature*—engage in a structured debate over multiple rounds (maximum five rounds in this work), and a dedicated *Judge* agent organizes their arguments into a final, user-facing report.

I. The *precursor agent* is prompted with domain knowledge and instructions focused on precursor selection. This agent can call predictive ML tools such as *ElemwiseRetro*,[13] and *SynthesisSimilarity*,[12] which propose candidate precursor sets and heating temperature. During the debate, the Precursor agent argues for synthesis routes that are compositionally plausible and experimentally accessible, grounded in the precursor ML tools.

II. The *thermodynamics agent* concentrates on bulk thermodynamics and phase stability. Its instructions emphasize concepts such as formation energy, phase competition, and phase diagrams. It is connected to bulk-level ML interatomic potential and structure-evaluation tools, including *MACE bulk*[47] and *MatterSim bulk*,[48] which are used to assess relative stability and energy landscapes of candidate structures and competing phases. Grounded in these tools, the Thermodynamics agent evaluates whether a proposed synthesis route is thermodynamically reasonable and flags routes that are likely to be unstable or lead to undesired phases at the target conditions.

III. The *surface & kinetics agent* focuses on surface energetics, diffusion, and kinetic accessibility. It has access to tools such as *MACE slab*, *MACE NEB*, *MatterSim slab*, *MatterSim NEB*, and *MechanismSurface*, which are used to estimate surface stability, reaction barriers, and diffusion or migration pathways relevant to the synthesis process (e.g., surface reactions, ion diffusion, or nucleation steps). In the debate, the Surface & Kinetics agent complements the Thermodynamics agent by highlighting kinetic bottlenecks, surface-mediated side reactions, or metastable but kinetically accessible routes, using the connected tools as evidence.

IV. The *literature agent* is responsible for grounding the debate in existing knowledge by performing *GPT-based web search* and *literature retrieval*. Its instructions emphasize fact-checking, comparison with reported synthesis routes, identification of similar chemistries, and extraction of qualitative mechanistic insight from prior work. During the debate, the Literature agent supports or challenges the proposals of the other agents by providing evidence from the published inorganic synthesis literature.

V. The *judge agent* finally aggregates the arguments produced by the four domain agents over multiple debate rounds. It is instructed to weigh thermodynamic plausibility, kinetic feasibility, precursor practicality, and literature consistency, and to produce a structured, human-readable output. The final report includes (i) a recommended synthesis method, (ii) a stepwise synthesis procedure, and (iii) a proposed mechanistic pathway.

Through this MAD process, *Materealize* (thinking mode) is able to deliver more refined, chemically grounded synthesis recommendations than those obtained by simply returning tool outputs. In typical use, thinking mode requires longer wall-clock time (~20 mins) than instant mode (1–2 mins), but in return, it provides substantially richer and more interpretable guidance on synthesis conditions and mechanistic pathways. We summarized the core functional modules and capabilities of *Materealize* in **Fig. 2**.

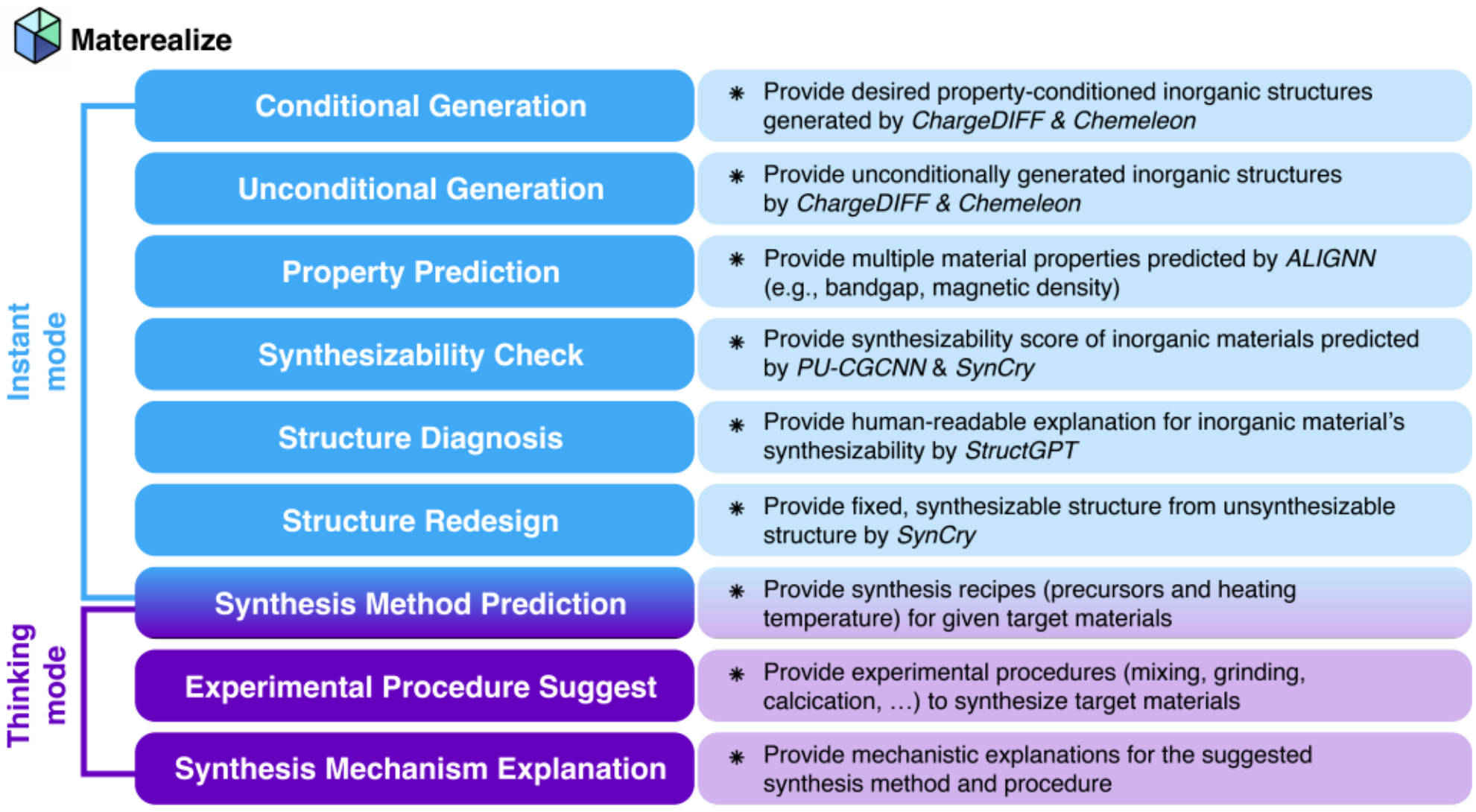

**Fig. 2 | Overview of *Materealize*'s core functional modules and capabilities.** In instant mode, *Materealize* can handle various tasks related to inorganic material design and synthesis within a few minutes. In thinking mode, *Materealize* can suggest more accurate and richer synthesis information within 20 minutes. We note that *Materealize* is designed to be highly extensible, making it easy to integrate new tools and seamlessly expand its functionality.

## Capabilities of Materealize across different prediction tasks.

To investigate the broad prediction task capabilities of *Materealize*, we applied it to four representative inorganic material-related tasks.

### Prediction Task 1: Conditional end-to-end design of synthesizable candidates

We first examined whether *Materealize* can propose synthesizable materials together with synthesis methods when a desired property is given. As an example, we set the target bandgap to 1.34 eV, near the Shockley-Queisser efficiency optimum,[49] which is desirable for perovskite solar cell application (**Fig. 3a**). Given the user query, "I want to find proper inorganic materials, which have bandgap around 1.34 eV.", *Materealize* interpreted the natural language query and used ChargeDIFF to perform property-conditioned structure generation with bandgap = 1.34 eV, generating 10 candidate structures. After that, the agent evaluated their synthesizability using PU-CGCNN and SynCry, and automatically filtered out candidates predicted to be unsynthesizable (synthesizability score < 0.5). Finally, *Materealize* successfully applied ElemwiseRetro to the remaining synthesizable candidates (e.g., $KCa_2O_2$, $Ti(FeO_2)_3$, …) to predict precursor sets and heating temperatures (e.g., For $KCa_2O_2$, Method 1 - Precursors: $K_2CO_3$+$CaCO_3$, Temperature: 885.1°C, Method 2 – Precursors: $K_2CO_3$+CaO, Temperature: 921.5°C, …). Within 4 minutes, the user received four synthesizable candidates (structure.cif files) along with their corresponding synthesis methods.

### Prediction Task 2: Diagnosing and fixing unsynthesizable materials

We tested whether *Materealize* can diagnose why a given structure is unsynthesizable and suggest a synthesizable alternative (**Fig. 3b**). Given the user query, "I want to diagnose and fix this structure: [mp-541044.cif file path].", *Materealize* checked the synthesizability of the given structure using PU-CGCNN and SynCry (synthesizability score = 0.326). After that, the agent converted the structure into a textual description using the Robocrystallographer tool and fed it to StructGPT to obtain explanations of why the structure is predicted to be unsynthesizable. After that, *Materealize* called SynCry to propose a fixed, more synthesizable structure, and re-evaluated it with PU-CGCNN and SynCry (fixed score = 0.622 > 0.5), then saved the redesigned structure as a synthesizable alternative. The process was completed in about one minute.

**Prediction Task 3: Detailed synthesis procedure and mechanism prediction**

To demonstrate *Materealize*'s thinking mode, we tested whether *Materealize* can suggest a detailed, experimentally actionable synthesis procedure together with a plausible reaction mechanism for either (i) synthesizable candidates produced by *Materealize* (e.g., "candidate 3") or (ii) a user-provided crystal structure (**Fig. 3c**). Given the user query, "Please generate a detailed synthesis procedure and possible reaction mechanism for candidate 3" (or "…for this structure: [CIF file path]"), *Materealize* initiated the corresponding multi-agent synthesis reasoning workflow. In this mode, the 4 agents (precursor, thermodynamics, surface & kinetics, literature) iteratively refined the plan over up to five rounds (precursor selection → reaction mechanism → thermodynamic conditions → step-by-step protocol → final consensus), minimizing repetition while explicitly tracking convergence via an agreement score. Finally, a Judge agent consolidated the 4 agent-level evidence into a single recommendation and produced a condensed human-readable synthesis report, including (1) precursors and stoichiometry, (2) mixing/mechanical activation (e.g., milling guidance), (3) stepwise heat-treatment schedules, (4) atmosphere/redox control as a key lever, and (5) a mechanism narrative describing intermediate formation and the rate-limiting stage (e.g., bulk cation diffusion during high-temperature conversion). We note that this report is not generated solely by LLM reasoning; instead, it is grounded in distinct tools, making it more reliable. **Fig. 4** illustrates an example dialogue trace of this multi-agent debate, demonstrating how the initial precursors suggested by ElemwiseRetro are iteratively refined through agent-level critique and ultimately finalized into a synthesis recommendation by the Judge agent. In a typical run, *Materealize* completed this end-to-end deliberation and produced the final report within 20 minutes.

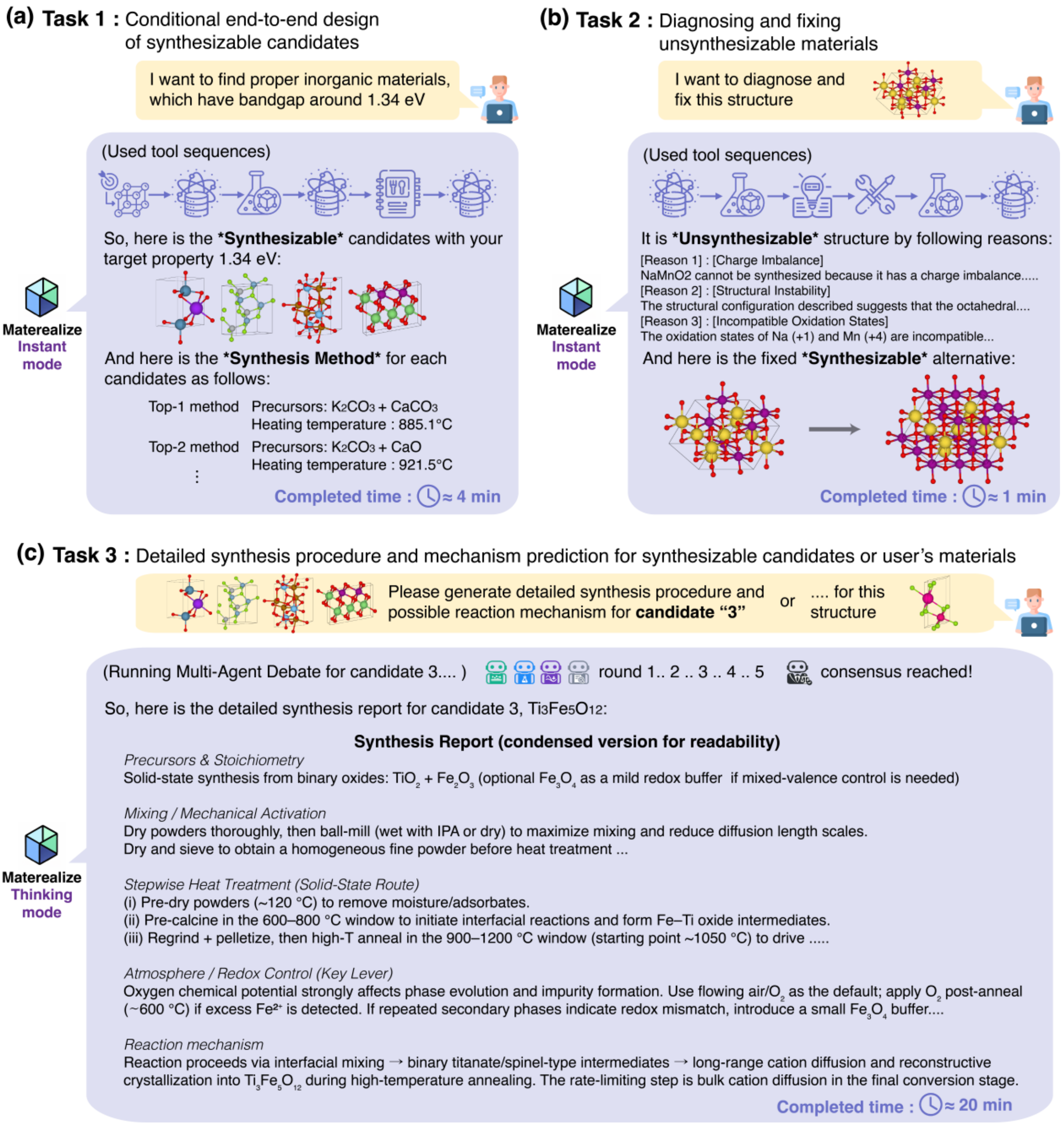

**Fig. 3 | Example of task capabilities by *Materealize*.** (a) Conditional end-to-end design of synthesizable candidates by instant mode. (b) Diagnosing and fixing unsynthesizable materials by instant mode. (c) Generating detailed synthesis reports for candidates from task (a) or starting with new materials by thinking mode.

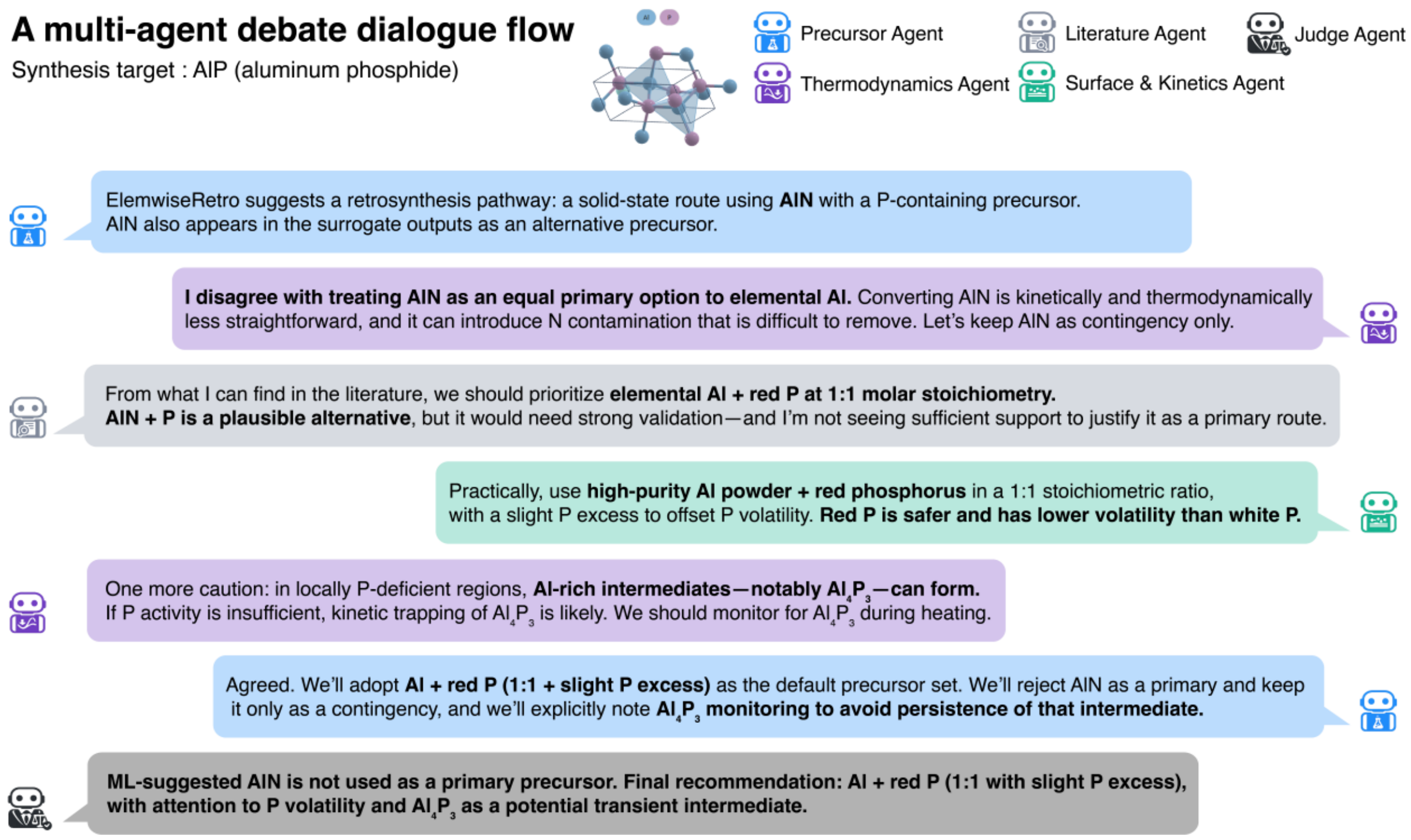

**Fig. 4 | Example of multi-agent debate dialogue trace by *Materealize* thinking mode.** Given a synthesis target, *Materealize* starts from the precursor set proposed by the initial tool (ElemwiseRetro) and then conducts a structured debate in which specialized agents (precursor, thermodynamics, literature, and surface/kinetics) iteratively critique and refine the route, ultimately converging on an improved synthesis recommendation.

### Predictive Task 4: Unconditional synthesizable material data augmentation

Additionally, we examined whether *Materealize* can be used for unconditional generation and collection of synthesizable inorganic materials. Given the user query, "I want to collect synthesizable material data.", *Materealize* used ChargeDIFF in unconditional mode to generate random inorganic crystal structures, and evaluated the synthesizability of all generated structures with PU-CGCNN and SynCry. Finally, the agent used ElemwiseRetro for adding each synthesis method and automatically saved them, thereby constructing a dataset of synthesizable inorganic materials for later use.

## Materealize performance benchmark

To evaluate *Materealize* performance on key tasks—including structure generation and synthesis recipe prediction—we compared five agent systems: *Materealize* (with connected tools), and four tool-free agent systems (gpt-4o, gpt-4o-mini, gemini-2.5-flash, and gemini-2.5-flash-lite) (**Fig. 5**).

For the structure generation benchmark, we performed 400 total generations (100 unconditional, 100 bandgap-conditioned, 100 magnetic density-conditioned, and 100 crystal density-conditioned). For the generated structures, we evaluated: (i) CIF-parsing validity (whether the CIF can be correctly loaded/saved), (ii) formula SMACT validity (whether the formula is valid by SMACT[50]) (iii) interatomic-distance validity (whether any minimum interatomic distance is > 0.5 Å), (iv) oxidation state validity (whether the oxidation state is valid by Pymatgen[45]) and (v) thermodynamic stability validity (whether CHGNet[51] v.0.3.0-calculated energy above hull is ≤ 0.2 eV/atom relative to MP references). Overall, *Materealize* showed a higher fraction of valid structures because it relies on a well-trained generative tool (**Fig. 5a-e**). In contrast, the tool-free agent systems (gpt-4o, gpt-4o-mini, gemini-2.5-flash, and gemini-2.5-flash-lite) produced structures that were rarely thermodynamically stable, underscoring the need for domain-specific tools for structure generation.

For the synthesis recipe benchmark, we also evaluated precursor prediction using a hold-out test set not used to train the connected tool. *Materealize* achieved higher top-k accuracy (k = 1, 2, …, 5) than the tool-free agent systems across all k values (**Fig. 5f**), again supporting the importance of integrating domain-trained tools for realistic synthesis planning. Especially, with the introduction of *Materealize* (thinking mode), which is grounded in tool outputs and performs multi-round debate, precursor prediction accuracy improved substantially beyond both instant mode and the tool-free baselines, achieving 86.2%, 90.2%, and 91.4% for top-3, top-4, and top-5 accuracy, respectively. Together, these results suggest that instant mode offers a strong baseline at low latency, whereas tool-grounded reasoning and multi-agent debate are essential to substantially improve top-k performance, resulting in top-5 accuracy exceeding 91% on this benchmark.

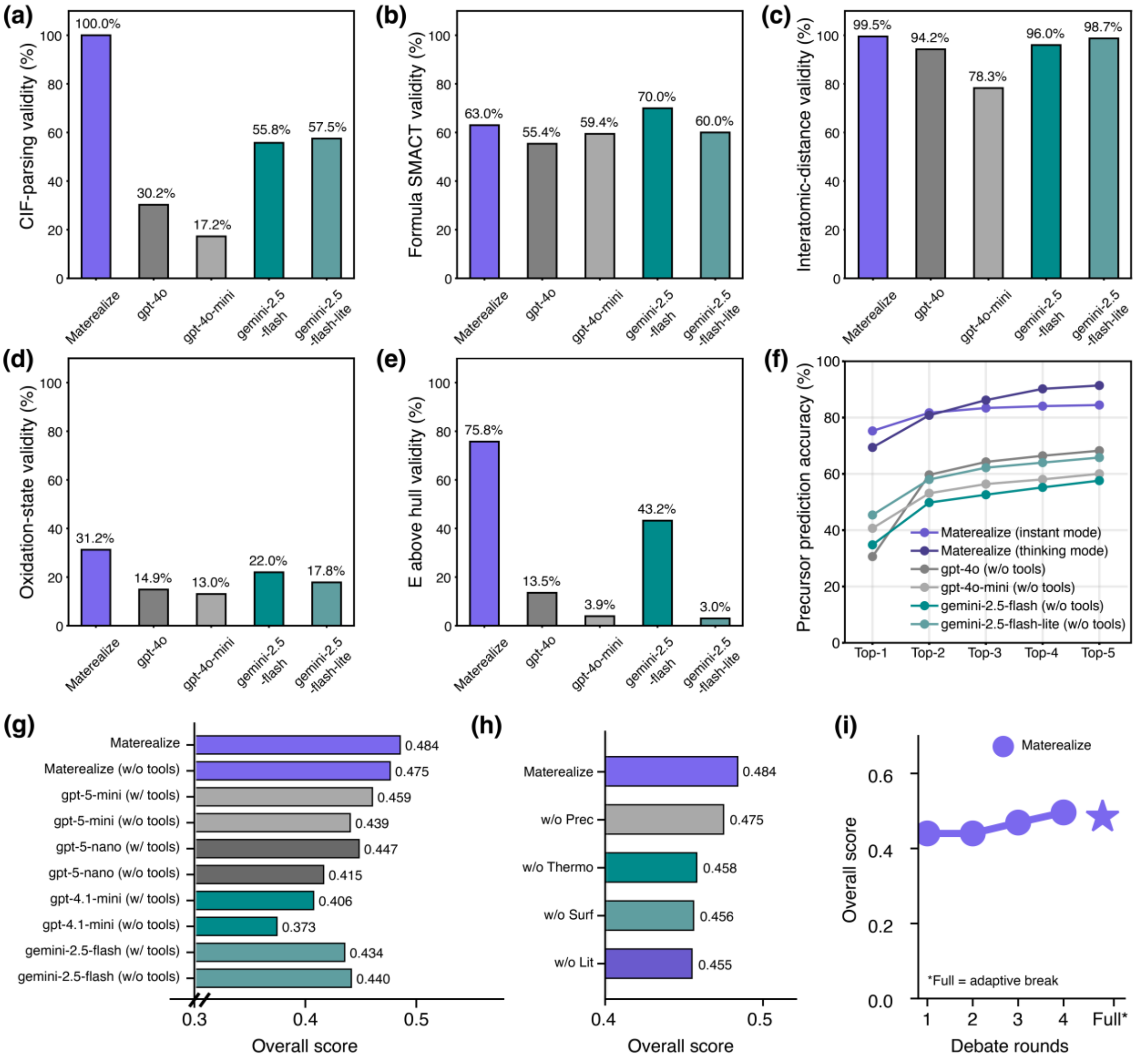

**Fig. 5 | Agent performance benchmark.** (a-e) Generated structure validity for (a) CIF-parsing, (b) formula SMACT check, (c) minimum interatomic-distance > 0.5 Å, (d) oxidation-state, and (e) energy above hull ≤ 0.2 eV/atom relative to MP references. (See also **Extended Data Fig. 1** and **2** for element diversity and energy above hull distribution.) (f) Top-k exact match accuracy of suggested precursors. (g-i) Generated synthesis method and mechanism evaluation.

To benchmark *Materealize*'s thinking mode performance for predicting synthesis method and mechanism, we evaluated its generated synthesis reports using a separate LLM-based evaluator[52,53] (**Extended Data Fig. 3**). We curated 20 mechanism-focused synthesis case studies, where each case consists of a full-text experimental paper and its corresponding target crystal structure (CIF). For each paper, we first used GPT-5 native PDF reading to identify the target compound and generate an "Answer report" that summarizes experimentally supported synthesis information across the four axes (Precursor, Temperature, Procedure, Mechanism). We then executed multiple model configurations (including *Materealize*) to produce "Test reports" from the same input CIF (and, when applicable, the same tools usage), and finally compared each Test report against the Answer report using an LLM-based evaluator. To reduce stochasticity in LLM judging, each configuration was evaluated 10 times per case, and we report the mean and standard deviation across runs. The full numerical results are illustrated in **Extended Data Fig. 4**.

We deeply examined *Materealize* against several single-model baselines under two settings: (i) tool-free reasoning using only the CIF and prompt, and (ii) tool-connected baselines that receive the same tool results but do not perform multi-agent debate. The LLM-evaluator assigns a normalized score for each axis (precursor, temperature, procedure, and mechanism) with an equal-weight overall score defined as their average. Across the 20-paper benchmark, *Materealize* achieved the strongest overall performance among the reported systems (**Fig. 5g**), reaching 0.484 overall score and outperforming tool-connected single-model baselines such as gpt-5-mini (w/ tools) (0.459) as well as tool-free baselines including gemini-2.5-flash (w/o tools) (0.440) and gpt-4.1-mini (w/o tools) (0.373). Furthermore, *Materealize* outperforms *Materealize* (w/o tools) (0.475), demonstrating that tool-grounded debate and reasoning are more effective than tool-free debate. Collectively, these results show that tool-grounded multi-agent deliberation with debate is more effective than both (i) debate-only reasoning without tools and (ii) single-agent reasoning that uses tools without multi-agent critique, highlighting the value of combining tool grounding with a role-specialized debate system.

We further conducted controlled ablation studies to quantify the contribution of agent roles and debate depth (**Fig. 5h**, **i**). Agent ablation results showed that removing any agent led to a drop in performance, with scores decreasing to 0.475, 0.458, 0.456, and 0.455 when the Precursor, Thermodynamics, Surface/Kinetics, and Literature agents were excluded, respectively, highlighting that synthesis reports really benefit from multi-perspective constraints rather than a single dominant heuristic. Additionally, increasing the number of debate rounds improved performance up to a plateau: *Materealize* increased from 0.440 (1 round) to 0.469 (3 rounds) and peaked at 0.496 (4 rounds), compared with the adaptive-break–enabled baseline (0.484). These results suggest that the current setting is a reasonable operating point, capturing most of the benefit from additional deliberation while avoiding diminishing returns.

Overall, the thinking mode benchmark demonstrated that multi-perspective debates, coupled with domain tool grounding, enable more reliable synthesis-condition and mechanism descriptions from crystallographic inputs.

## Simulation for synthesis report validation

While we quantified *Materealize*'s performance using benchmark metrics and LLM-based evaluations (**Fig. 5**), we additionally performed a physics-grounded validation to evaluate whether the *Materealize*-suggested synthesis methods led to physically reasonable precursor-to-product transformations under the proposed conditions. Accordingly, we conducted MACE-based machine-learning-potential molecular dynamics (MLP–MD) simulation for novel $Mg_2TeSe$ target. We built atomistic models of the *Materealize*-specified precursors, arranged them in a stacked configuration to mimic solid-solid contact, and followed the *Materealize*-suggested thermal profile using a ramp-to-peak step and an isothermal anneal. Because experimental synthesis occurs on much longer timescales and at substantially larger spatial scales than are accessible to atomistic simulations, we used an accelerated protocol while preserving the key causal factors—precursor selection, stoichiometry, and peak temperature. We then computed powder XRD patterns from the final simulated structure and compared it with reference peak derived from the *Materealize*-provided target crystal structures (**Fig. 6**). As a result, the simulated pattern reproduced the dominant Bragg features of the intended phases, with trajectories showing qualitative mixing and ordering consistent with high-temperature solid-state reactions. We also illustrated that the simulated structures contain crystal structures highly similar to those of the *Materealize*-generated material. Taken together, these observations support that *Materealize*-generated novel materials and synthesis methods are chemically plausible by the proposed precursor and thermal conditions.

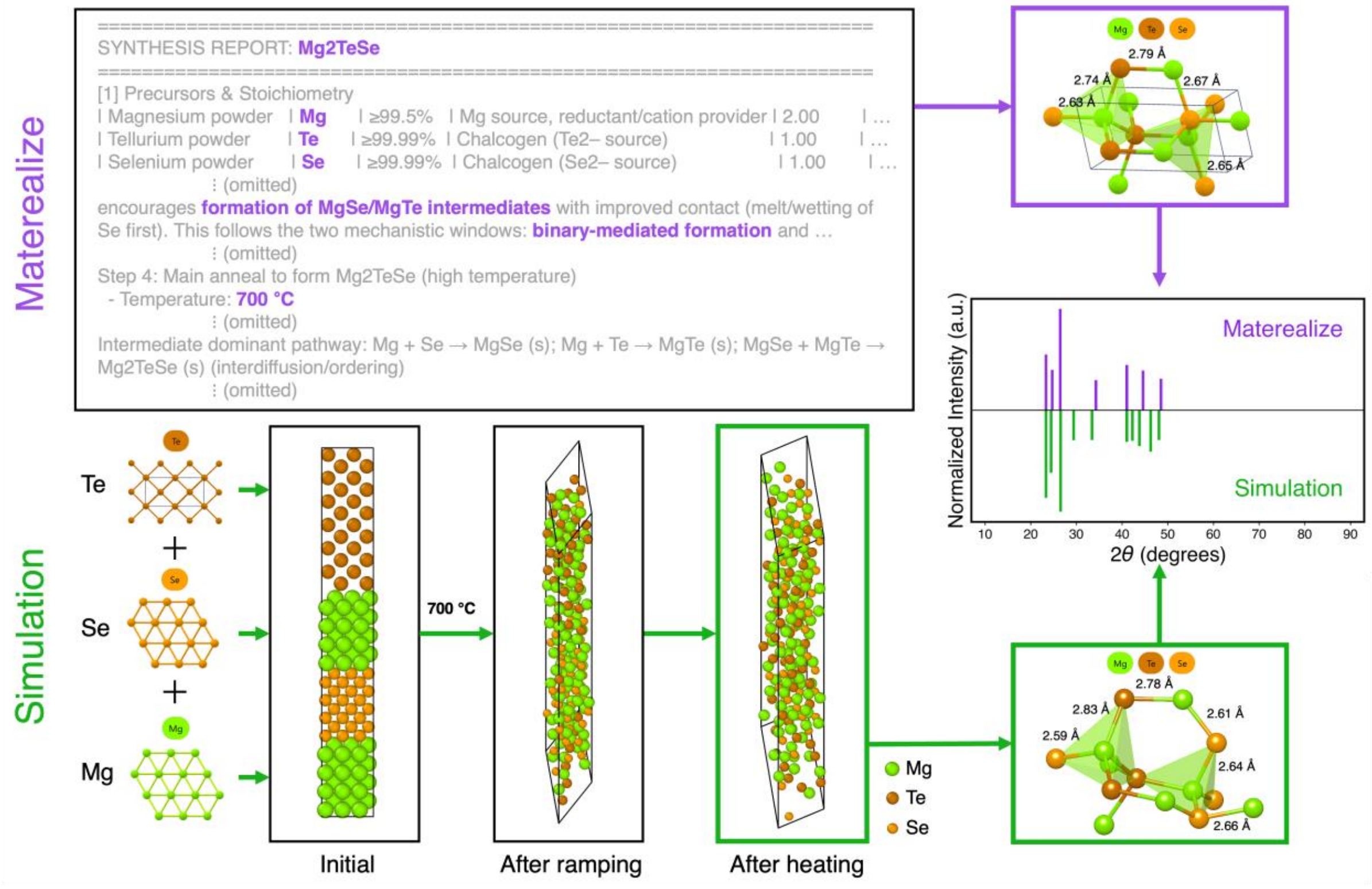

**Fig. 6 | Simulation-based validation of *Materealize* synthesis report.** For the $Mg_2TeSe$ target, *Materealize* generates a synthesis report and the corresponding target structure. We convert each report into a MACE-based molecular dynamics (MD) validation workflow by building an initial stacked precursor assembly (Initial), applying temperature ramping to the report-specified (700 °C) peak temperature (After ramping), and conducting isothermal annealing (After heating). We then compare powder XRD computed from the relaxed structure obtained by further optimizing the final MD snapshot (green) with the reference peaks from the target structure (purple). The simulated patterns reproduce the dominant target reflections.

## Extending the synthesis-realizable design space

We quantitatively analyzed how much the synthesis-realizable design space is extended by *Materealize*, particularly in thinking mode. We define the synthesis-realizable design space as:

$$\begin{aligned}&\text{Synthesis Realizable Design Space (\%)}\\&\qquad = (\text{Energetically stable ratio of generated structures; } E_{hull} \leq 0.2 \text{ eV/atom})\\&\qquad \times (\text{Valid synthesis method ratio within top 5 suggestions})\end{aligned}$$

Using this metric, we evaluated *Materealize* thinking mode and instant mode, and compared them with four tool-free agent systems (gpt-4o, gpt-4o-mini, gemini-2.5-flash, gemini-2.5-flash-lite). As shown in **Fig. 7a**, both *Materealize* instant and thinking modes achieve a substantially larger synthesis-realizable design space than tool-free agents, demonstrating that tool integration is critical for accurate inorganic design and synthesis. Notably, thinking mode outperforms instant mode by enabling an 8.3% broader designable region, which originates from its improved top-5 synthesis recipe prediction performance (**Fig. 5f**) enabled by tool-grounded multi-agent debate.

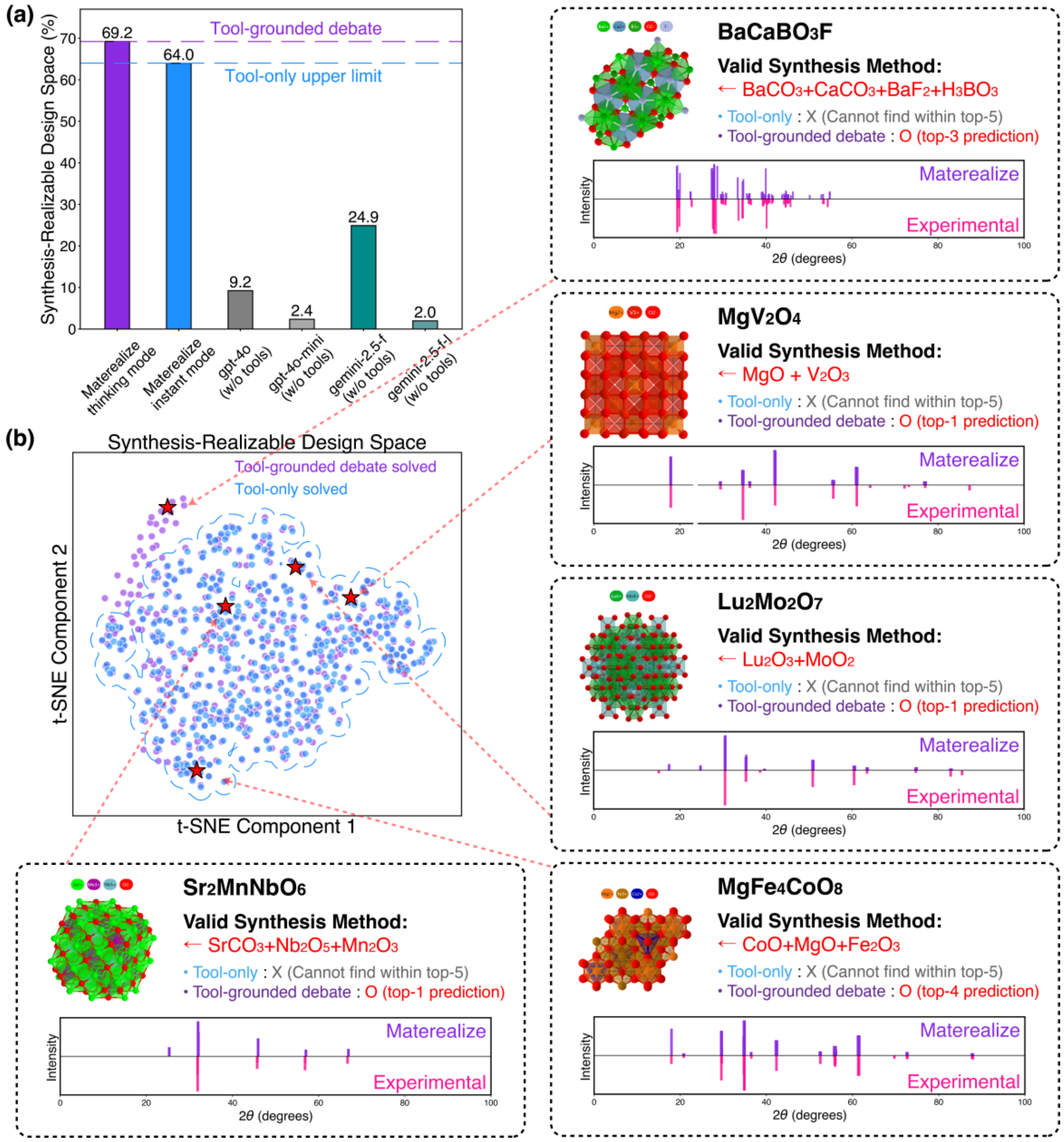

**Fig. 7 | Extending the synthesis-realizable design space with *Materealize*.** (a) Quantitative comparison of the synthesis-realizable design space across different agent systems. The synthesis-realizable design space is defined as the product of the thermodynamically stable ratio ($E_{hull} \leq 0.2$ eV/atom) and the valid synthesis-method ratio within the top-5 predictions. (b) t-SNE visualization of materials space for which correct design-and-synthesis routes are identified by *Materealize* in both instant and thinking modes. Among the materials designed by *Materealize*, five experimentally-validated materials are shown, where tool-only execution fails to identify valid synthesis routes within the top-5 predictions, while thinking mode successfully proposes experimentally consistent synthesis methods.

To further analyze where the design space expands, we visualized the material space using a t-SNE embedding of the correctly designed-and-synthesized materials from tool-only (instant mode) and tool-grounded debate (thinking mode) in **Fig. 7b**. The results show that thinking mode can successfully propose valid design-and-synthesis solutions beyond the boundary of tool-only execution, and also improves correctness within the tool-only domain. Among the several materials designed by *Materealize*, we found several real-world-validated materials ($BaCaBO_3F$, $MgV_2O_4$, $Lu_2Mo_2O_7$, $MgFe_4CoO_8$, and $Sr_2MnNbO_6$) for which the tool-only failed to identify a valid synthesis method within the top-5 suggestions, while the thinking mode successfully produced correct synthesis routes (**Fig. 7b**). This demonstrates that, compared to manually constructed fixed pipelines that sequentially connect several ML models, *Materealize* can further expand the feasible materials design-and-synthesis space through tool-grounded multi-agent debate.

## Public release of 100 synthesizable candidates with synthesis guidelines

We leveraged *Materealize* to propose a set of 100 synthesizable inorganic candidates conditionally generated with target bandgaps of 1.34 eV, 1.5 eV, and 2.0 eV (n = 51, 36, and 13, respectively), which are desirable properties for (i) perovskite solar cells, (ii) solar absorbers, and (iii) photocatalysts, respectively. As shown in **Fig. 8**, all proposed candidates were screened through synthesizability checks, ensuring a high likelihood of successful experimental realization. In addition, the candidates were selected to exhibit predicted properties closely matching the target values, thereby forming a set of materials that simultaneously satisfy both synthesizability and property-target criteria. Furthermore, for each candidate, *Materealize* provides detailed synthesis methods, including precursor combinations, processing conditions, and experimental procedures, enabling experimentalists to directly follow and test the proposed routes. We provide the complete list of all 100 promising synthesizable candidates in a public repository. Through this large-scale generation and synthesis-aware filtering, *Materealize* demonstrates its potential to accelerate practical materials realization by bridging computational design and experiment-ready synthesis guidance.

| Material | Designed information | Synthesis method |
|---|---|---|
| 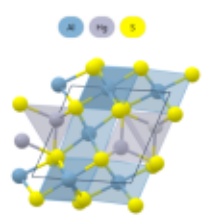 | Formula = $Al_2HgS_4$<br>Target Bandgap = 1.34 eV<br>Pred. Bandgap = 1.3418 eV<br>Synthesizability score = 73<br>Possible application:<br>- Perovskite solar cell | # For 5.00g batch of $Al_2HgS_4$,<br>$Al_2S_3$(1.96g)+HgS(3.04g)+S(optional)<br># Synthesis procedure<br>: Pre-mixing ⟶ Ampoule loading (60 min) ⟶ Low-temperature interdiffusion (400°C) ⟶ Crystallization annealing (550°C, 12h) ⟶ Optional post-anneal (300-400°C) ... |
| 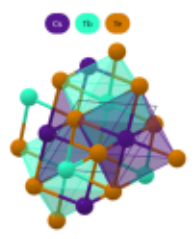 | Formula = $CsTbTe_2$<br>Target Bandgap = 1.34 eV<br>Pred. Bandgap = 1.3368 eV<br>Synthesizability score = 99<br>Possible application:<br>- Perovskite solar cell | # For 5.00g batch of $CsTbTe_2$,<br>$Cs_2CO_3$(1.49g)+Tb(1.45g)+Te(2.33g)<br># Synthesis procedure<br>: Drying ⟶ Initial mixing & pre-calcination (400°C, 8h) ⟶ Heating (750°C, 24h) ⟶ (Optional) Post-anneal (500°C, 48h) ⟶ ... |
| 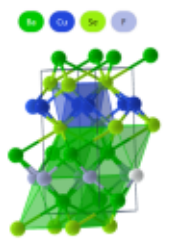 | Formula = BaCuSeF<br>Target Bandgap = 1.50 eV<br>Pred. Bandgap = 1.5061 eV<br>Synthesizability score = 99<br>Possible application:<br>- Solar absorber | # For 5.00g batch of BaCuSeF,<br>$BaF_2$(1.47g)+BaSe(1.81g)+Cu2Se(1.72g)<br># Synthesis procedure<br>: Pre-drying & mixing ⟶ Low-temperature activation (200°C, 4h) ⟶ Intermediate diffusion (500°C, 6h) ⟶ Crystallization (750°C, 24h) ⟶ ... |
| 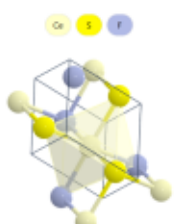 | Formula = $CeSF_2$<br>Target Bandgap = 1.50 eV<br>Pred. Bandgap = 1.4892 eV<br>Synthesizability score = 51<br>Possible application:<br>- Solar absorber | # For 5.00g batch of $CeSF_2$,<br>$CeF_3$(1.56g)+$Ce_2S_3$(2.99g)+$NH_4F$(0.88g)<br># Synthesis procedure<br>: Pre-drying & mixing ⟶ Cold pressing ⟶ Low-temperature decomposition (250°C, 2h) ⟶ ... |
| 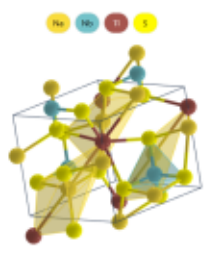 | Formula = $Na_2NbTlS_4$<br>Target Bandgap = 2.00 eV<br>Pred. Bandgap = 1.9826 eV<br>Synthesizability score = 85<br>Possible application:<br>- Photocatalyst | # For 5.00g batch of $Na_2NbTlS_4$,<br>$Nb_2O_5$(1.41g)+$Na_2S$(0.83g)+$Tl_2S$(2.34g)+Excess S(0.03g)<br># Synthesis procedure<br>: (In glovebox) Mixing ⟶ Encapsulation (sealed ampoule) ⟶ Sulfidation (350°C, 8h) ⟶ ... |

**Fig. 8 | Examples of synthesizable inorganic candidates designed by *Materealize*.** The synthesis methods are summarized in an abbreviated format (see **Fig. 3c** for improved readability). The complete list of all 100 promising synthesizable candidates with detailed synthesis procedures is provided in a public repository. Alternatively, you can also view the full set of 100 entries by visiting our demo web app and navigating through the interface (https://huggingface.co/spaces/SNU-MICC/Materealize).

## Conclusions

In this work, we introduced *Materealize*, a unified agentic framework for end-to-end inorganic materials design and synthesis planning. Compared to existing inorganic design agents,[33-36] *Materealize* goes beyond inorganic materials design by also addressing accurate synthesis planning and mechanism suggestion, thereby providing more actionable guidance for experimentalists. *Materealize* provides two complementary operation modes: an instant mode that rapidly composes domain tools for property-conditioned generation, synthesizability assessment, and synthesis recipe prediction, and a thinking mode that performs tool-grounded multi-agent debate (MAD) for synthesis procedures and mechanistic hypotheses. By automating tool orchestration through a natural-language interface, *Materealize* lowers the barrier for non-experts while enabling expert-level, synthesis-aware investigation.

We systematically evaluated *Materealize* across representative inorganic tasks and controlled benchmarks. In instant mode, *Materealize* achieved higher validity and practicality in generated structures and synthesis suggestions by leveraging well-trained domain tools, whereas tool-free agent baselines frequently produced outputs that violated basic physical or thermodynamic constraints. In thinking mode, *Materealize*'s role-specialized agents collaboratively critique, cross-check, and refine heterogeneous tool-derived hypotheses, producing synthesis reports that are more accurate and chemically plausible than both (i) debate-only reasoning without tools and (ii) single-agent tool-connected reasoning without multi-agent critique. We further performed a physics-grounded validation using MACE-based MLP–MD, showing that *Materealize*-reported precursor/thermal conditions yield annealed structures whose powder XRD reproduces the dominant Bragg features of the intended targets. Furthermore, *Materealize* expands the synthesis-realizable design space by 8.3% beyond manually constructed fixed ML pipelines for inorganic design and synthesis, highlighting the advantage of tool-grounded multi-agent debate. We provide 100 synthesizable candidates with detailed synthesis guidance, to accelerate experiment-ready inorganic materials discovery by bridging computational design and practical realization.

However, although we used the multi-agentic debate approach to go beyond single-tool outputs via cross-checking, it still fundamentally depends on the coverage and accuracy of the tools it is connected to. Importantly, many underlying models are trained on currently available materials databases[54-57] that are dominated by positive (successful) examples and contain limited negative or disordered data, which can lead to selection bias.[58,59] Future progress will therefore require unbiased, multi-variate data collection that includes negative outcomes, potentially enabled by emerging self-driving laboratories (SDLs)[60,61] and closed-loop active learning.[62] Additionally, tighter integration with higher-fidelity verification (e.g., first-principles calculations and MD simulations) and calibrated uncertainty reporting will further strengthen reliability for high-stakes experimental decision-making.

## Methods

### Agent system construction

The LLM used in *Materealize* instant mode is based on *OpenAI*'s *gpt-4o model*, with tool integration and LLM invocation managed through the *Pydantic AI package* (v.1.16.0). A total of 12 machine learning model endpoints are connected: 5 generative models (*ChargeDIFF* & *Chemeleon* for unconditional, bandgap-conditional, magnetic density-conditional, crystal density, composition-conditional generations), 1 property predictive model (*ALIGNN*), 3 synthesizability prediction models (*SynthGPT* for compositional model, and *SynCry* & *PU-CGCNN* for structural model), synthesizability explanation model (*StructGPT*), structure redesign model (*SynCry*), and synthesis method prediction model (*ElemwiseRetro*). For handling inorganic material structure files (CIF files), the system uses *Pymatgen* (v.2025.10.7) built-in functions for loading and saving. Conversion to crystal graph representation follows the CGCNN method, conversion to CIF-string representation follows the *SynCry* method, and conversion to descriptive textual representation is obtained by *Robocrystallographer* (v.0.2.11) tool. These inorganic data conversion and I/O processes are all integrated into the *Materealize*, allowing the agent to perform them automatically. Inputs and outputs within the agent are structured using *Pydantic*'s *BaseModel* instead of a text-based calling system, eliminating the need for retry logic and preventing unexpected response blocks.

### Evaluation

The agent system was evaluated by following methods. For structure generation performance, we checked CIF-parsing validity, formula SMACT validity, interatomic-distance validity, oxidation-state validity, and energy above hull validity, as described in the main-text. For synthesis recipe prediction performance, we evaluated the top-k exact match accuracy for the hold-out-test set from the text-mined inorganic synthesis dataset.[57] For evaluating synthesis procedures and mechanisms, we collected 20 case studies with experimentally elucidated mechanisms and used a separate LLM-based evaluator to score the outputs by comparing the "Answer report" against the "Test report", as described in the main-text.

### Simulation details

All simulations were performed using the MACE-MP-0b3-medium[47] machine-learned interatomic potential under the 3D periodic boundary conditions with the Atomic Simulation Environment (ASE).[63]

Initial bulk precursor structures were obtained from the Materials Project[54] (MP) database and expanded to supercells for the solid-state reaction simulations. Heterointerface models were generated by stacking slabs exposing low-index surface orientations with an initial interfacial spacing of ~3.0 Å to avoid unphysical atomic overlap. In-plane supercell matches were identified using the Zur–McGill lattice-matching algorithm,[45] and the heterointerface model requiring the least in-plane strain was selected as the initial geometry.

The initial heterointerface structure was relaxed using the BFGS algorithm with a maximum force convergence criterion ($f_{max}$) of 0.01 eV Å$^{-1}$. After relaxation, NPT molecular dynamics simulations using the Nosé–Hoover thermostat and Parrinello–Rahman barostat were performed with a fully flexible triclinic cell system to raise the temperature up to the target during 100 ps. Subsequently, the NPT simulation was performed at the target temperature for over 0.5 ns to ensure that the solid-state reaction occurred sufficiently. Starting from the final MD snapshots, we performed a post-MD structural relaxation until the $f_{max}$ was less than 0.001 eV Å$^{-1}$, and subsequently computed powder XRD patterns from the relaxed structures. To account for peak broadening effects, the resulting stick patterns were Gaussian-broadened via convolution using a standard deviation of 0.3 degrees in 2θ. To compare XRD peaks with the *Materealize*-generated reference structures, we extracted the main peaks from the raw simulated XRD data using the *Peakutils* Python package. The resulting structures were visualized using OVITO[64] and VESTA.[65] The $Mg_2TeSe$ heterointerface was modeled by stacking Mg (mp-110) (100), Se (mp-119) (100), and Te (mp-10654) (010), and the final NPT production was performed at 700 °C for 1.8 ns.

## Code availability

The code underlying this study is available upon reasonable request. A demonstration web interface for *Materealize* is available at https://huggingface.co/spaces/SNU-MICC/Materealize. The public release of 100 synthesizable candidates with synthesis guidelines is available at https://zenodo.org/records/18240693.

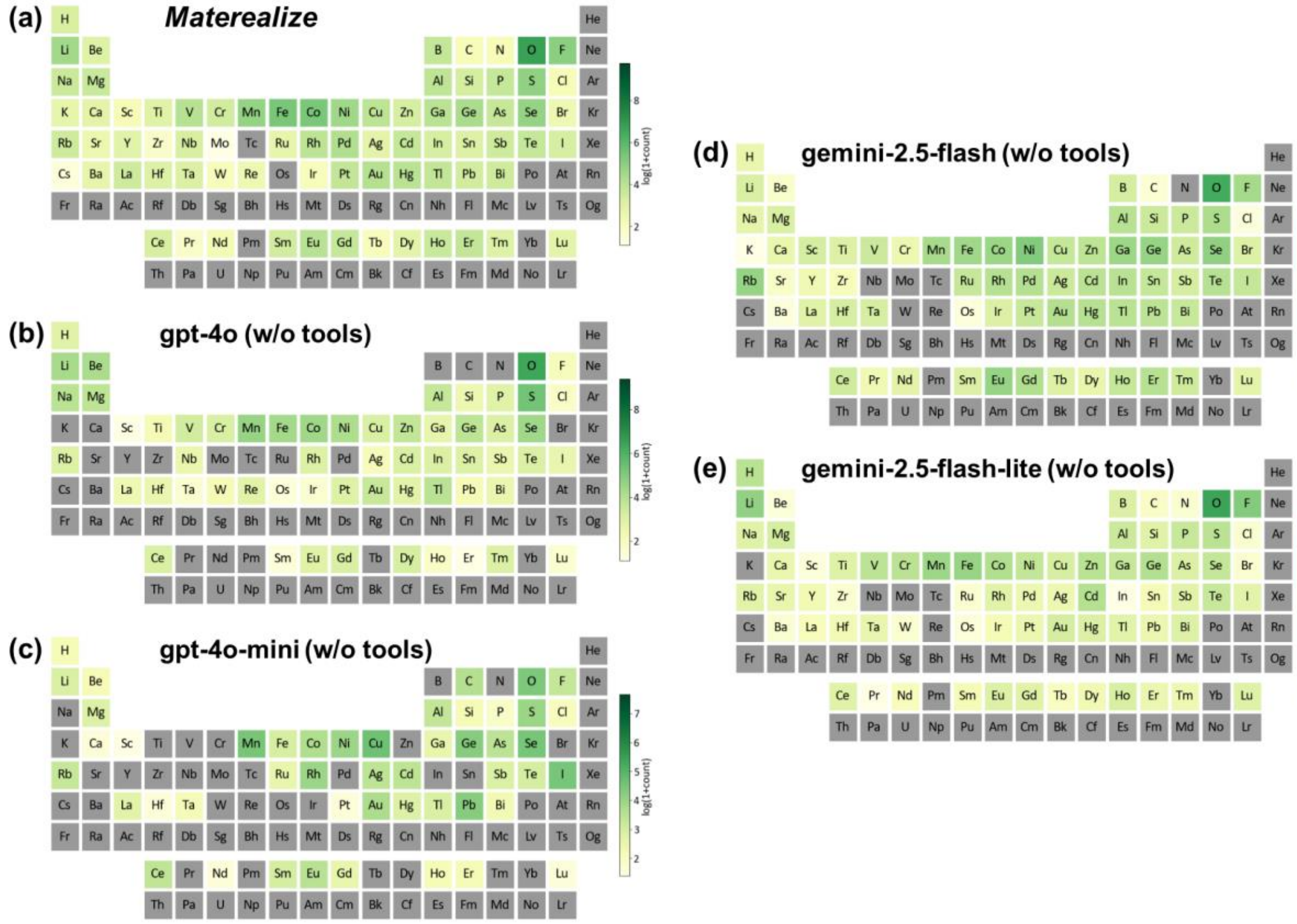

**Extended Data Fig. 1 | Element distributions of generated structures** by (a) *Materealize*, (b) gpt-4o (w/o tools), (c) gpt-4o-mini (w/o tools), (d) gemini-2.5-flash (w/o tools), and (e) gemini-2.5-flash-lite (w/o tools). These results demonstrate that the tool-connected *Materealize* is able to span the broadest range of chemical systems than the tool-free agents (gpt-4p, gpt-4o-mini, gemini-2.5-flash, and gemini-2.5-flash-lite).

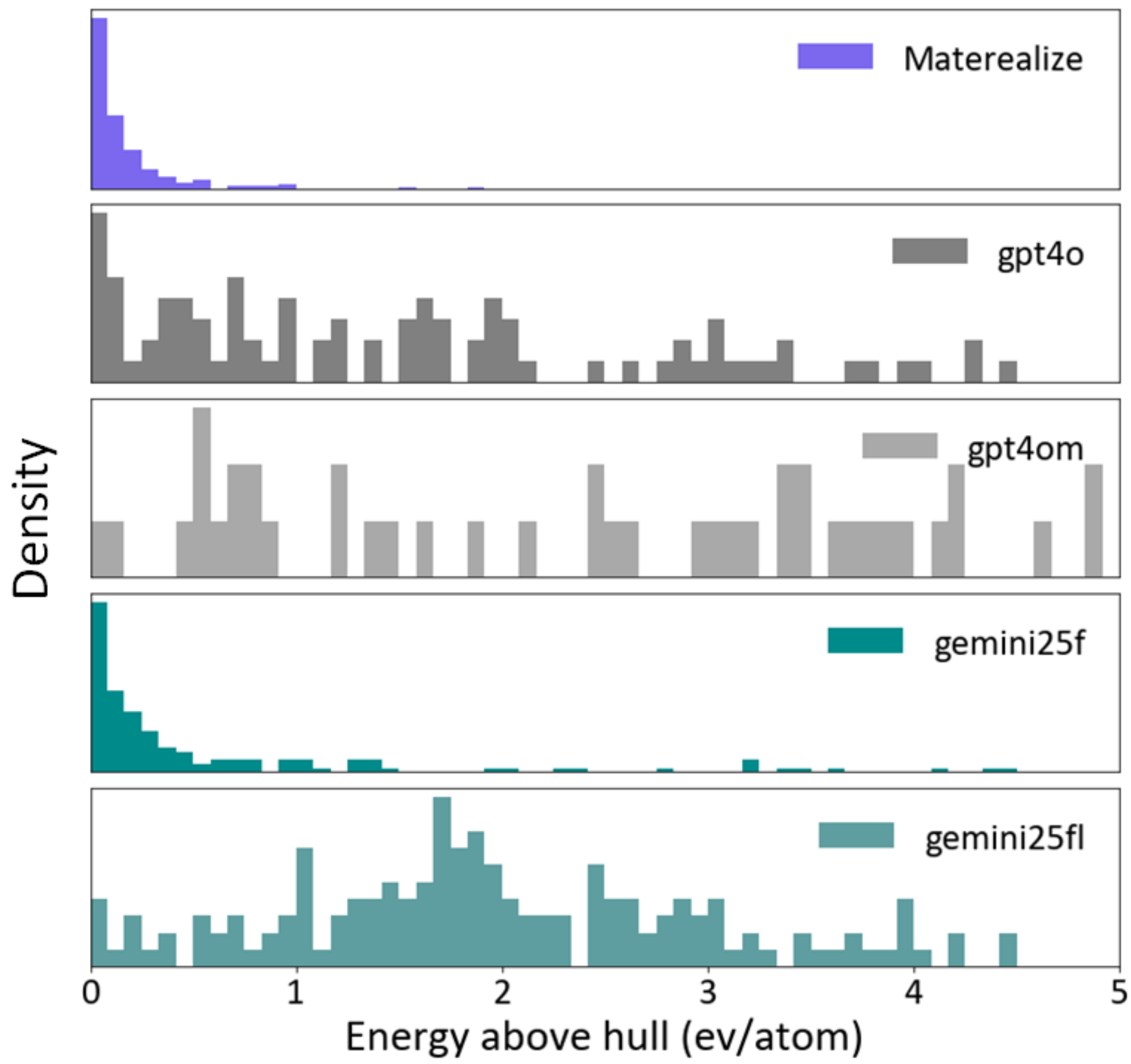

**Extended Data Fig. 2 | Energy above hull distributions of generated structures** by *Materealize*, gpt-4o (w/o tools), gpt-4o-mini (w/o tools), gemini-2.5-flash (w/o tools), and gemini-2.5-flash-lite (w/o tools). These results indicate that the tool-connected *Materealize* is able to reliably generate thermodynamically stable structures.

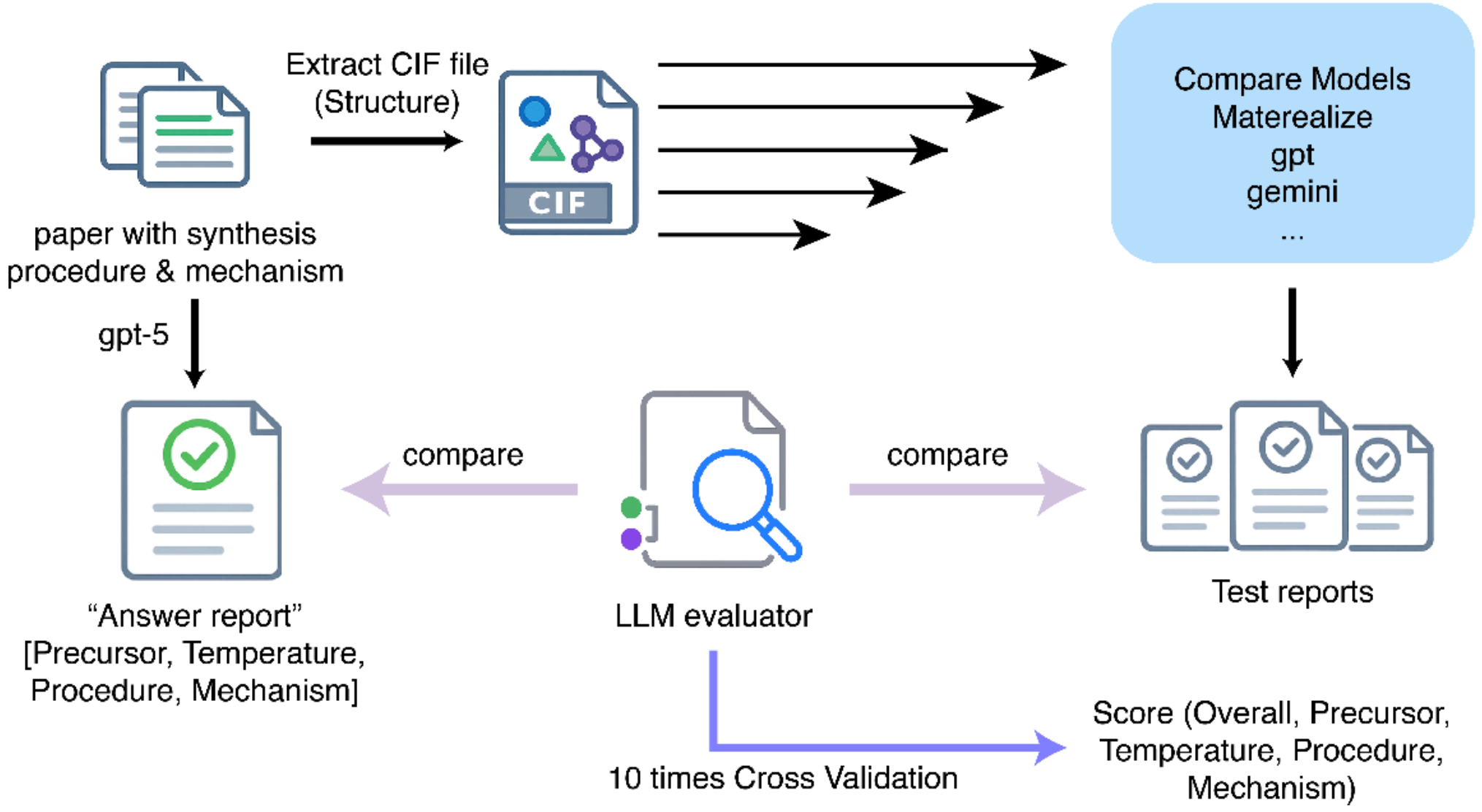

**Extended Data Fig. 3 | LLM-evaluator framework for synthesis-report benchmark.** For each literature case, we extract the target structure (CIF) and the synthesis description, and compile a reference "answer report" (precursor, temperature, procedure, mechanism). Multiple models then generate corresponding test reports from the same CIF. An LLM-based evaluator compares each test report to the answer report and assigns an overall score along with category-wise scores (precursor, temperature, procedure, mechanism), aggregated over 10× cross-validation runs.

| Method | Overall | Precursor | Temp | Procedure | Mechanism |
|---|---|---|---|---|---|
| Materealize (4 rounds) | 0.496 ± 0.195 | 0.533 ± 0.242 | 0.509 ± 0.165 | 0.450 ± 0.192 | 0.486 ± 0.210 |
| Materealize | 0.484 ± 0.169 | 0.536 ± 0.217 | 0.474 ± 0.153 | 0.435 ± 0.166 | 0.486 ± 0.171 |
| Materealize (– Precursor Agent) | 0.475 ± 0.170 | 0.502 ± 0.219 | 0.507 ± 0.164 | 0.431 ± 0.161 | 0.458 ± 0.175 |
| Materealize (w/o tools) | 0.475 ± 0.188 | 0.521 ± 0.249 | 0.477 ± 0.164 | 0.432 ± 0.184 | 0.468 ± 0.193 |
| Materealize (3 rounds) | 0.469 ± 0.186 | 0.518 ± 0.232 | 0.477 ± 0.177 | 0.424 ± 0.185 | 0.457 ± 0.181 |
| GPT-5-mini (w/ tools) | 0.459 ± 0.162 | 0.532 ± 0.212 | 0.479 ± 0.160 | 0.416 ± 0.177 | 0.407 ± 0.156 |
| Materealize (– Thermodynamics Agent) | 0.458 ± 0.186 | 0.497 ± 0.232 | 0.477 ± 0.173 | 0.413 ± 0.179 | 0.442 ± 0.186 |
| Materealize (– Surface/Kinetics Agent) | 0.456 ± 0.201 | 0.490 ± 0.245 | 0.460 ± 0.186 | 0.411 ± 0.201 | 0.459 ± 0.190 |
| Materealize (– Literature Agent) | 0.455 ± 0.193 | 0.513 ± 0.241 | 0.459 ± 0.176 | 0.408 ± 0.190 | 0.440 ± 0.198 |
| GPT-5-nano (w/ tools) | 0.447 ± 0.141 | 0.532 ± 0.205 | 0.439 ± 0.127 | 0.406 ± 0.143 | 0.406 ± 0.126 |
| Materealize (2 rounds) | 0.440 ± 0.198 | 0.482 ± 0.223 | 0.447 ± 0.196 | 0.404 ± 0.199 | 0.424 ± 0.196 |
| Materealize (1 round) | 0.440 ± 0.179 | 0.500 ± 0.233 | 0.454 ± 0.180 | 0.395 ± 0.177 | 0.410 ± 0.168 |
| Gemini 2.5 Flash (w/o tools) | 0.440 ± 0.147 | 0.505 ± 0.210 | 0.433 ± 0.134 | 0.395 ± 0.146 | 0.424 ± 0.164 |
| GPT-5-mini (w/o tools) | 0.439 ± 0.198 | 0.513 ± 0.250 | 0.430 ± 0.193 | 0.398 ± 0.200 | 0.411 ± 0.185 |
| Gemini 2.5 Flash (w/ tools) | 0.434 ± 0.174 | 0.516 ± 0.233 | 0.428 ± 0.155 | 0.389 ± 0.176 | 0.398 ± 0.172 |
| GPT-5-nano (w/o tools) | 0.415 ± 0.178 | 0.490 ± 0.238 | 0.418 ± 0.153 | 0.378 ± 0.173 | 0.372 ± 0.179 |
| GPT-4.1-mini (w/ tools) | 0.406 ± 0.145 | 0.471 ± 0.180 | 0.438 ± 0.166 | 0.359 ± 0.141 | 0.355 ± 0.137 |
| GPT-4.1-mini (w/o tools) | 0.373 ± 0.139 | 0.441 ± 0.194 | 0.390 ± 0.128 | 0.339 ± 0.140 | 0.319 ± 0.121 |

**Extended Data Fig. 4 | Synthesis report validation results(all) sorted by rank.** Performance comparison of *Materealize* and baseline LLMs on the synthesis-report benchmark. Methods are ranked by the overall LLM-evaluator score, and scores are reported as mean ± standard deviation over repeated evaluation runs. Category-wise scores quantify model accuracy for precursor selection, temperature profile, procedure description, and mechanistic explanation. "w/ tools" indicates tool-connected inference, while *Materealize* variants (e.g., different rounds or with specific agents removed) assess the contribution of individual components to overall synthesis-report quality.